%% file: 2012_latcon-proc.tex
\RequirePackage{ifpdf}
\documentclass{PoS}
\usepackage{amsmath,graphicx}
\newcommand{\VmA}{V\!\!\!-\!A}
\DeclareMathOperator{\tr}{tr}
\DeclareMathOperator{\re}{Re}

\title{Lattice Hadron Structure: \\ Applications within and beyond QCD}
\ShortTitle{Lattice Hadron Structure: Applications within and beyond QCD}

\author{\speaker{Huey-Wen Lin}\footnote{NT@UW-12-19}
\\
Department of Physics, University of Washington, Seattle, WA 98195-1560 \\
}

\abstract{
Study of the hadronic matrix elements can provide not only tests of the QCD sector of the Standard Model (in comparing with existing experiments) but also reliable low-energy hadronic quantities applicable to a wide range of beyond-the-Standard Model scenarios where experiments or theoretical calculations are limited or difficult.
On the QCD side, progress has been made in the notoriously difficult problem of addressing gluonic structure inside the nucleon, reaching higher-$Q^2$ region of the form factors, and providing a complete picture of the proton spin. However, even further study and improvement of systematic uncertainties are needed.
There are also proposed calculations of higher-order operators in the neutron electric dipole moment Lagrangian, which would be useful when combined with effective theory to probe BSM.
Lattice isovector tensor and scalar charges can be combined with upcoming neutron beta-decay measurements of the Fierz interference term and neutrino asymmetry parameter to probe new interactions in the effective theory, revealing the scale of potential new TeV particles.
Finally, I revisit the systematic uncertainties in recent calculations of $g_A$ and review prospects for future calculations.
}

\FullConference{The XXX International Symposium on Lattice Field Theory\\
June 24--June 29, 2012\\
Cairns, Australia\\
}

\begin{document}

\vspace{-0.5cm}
\section{Introduction}
\vspace{-0.4cm}

\input{intro.tex}

\vspace{-0.5cm}
\section{Probing Inside Hadron}
\vspace{-0.3cm}
\input{struc.tex}

\vspace{-0.5cm}
\section{Applications beyond QCD}
\vspace{-0.3cm}
\input{bsm.tex}

\vspace{-0.5cm}
\section{Nucleon Axial Charge $g_A$}
\vspace{-0.3cm}
\input{gA.tex}

\vspace{-0.5cm}
\section{Summary and Outlook}
\vspace{-0.3cm}

There have recently been many new developments, calculations of new quantities and new ideas for exploring and extending the reach of lattice QCD for hadronic structure. Many of the calculations presented here may not be final numbers from lattice QCD that can be quoted by experiments or model builders, and they may not be completely free from uncontrolled systematics. Even so, such preliminary explorations represent exciting progress, and we need people to make aggressive calculations that push the frontier. Hadronic structure is essential to understanding how the quarks and gluons make up hadrons. We push their precision so that we can make tests of the Standard Model when compared with reliable experimental results or evaluate the validity of nonperturbative QCD models. Hadronic low-energy couplings can be contributed to effective theories to extend the reach of theory to processes that are too complicated to directly calculate on the lattice, and they can be used in combination with precision low-energy experiments to probe new interactions induced by potential few-TeV BSM particles. We have moved from a period where we barely understood potential systematics to an era of examining them with greater statistics. By the application of increasing computational resources, we will push to larger volumes ($M_\pi L \geq 6$) at physical pion mass and longer ensemble trajectories so that we will be able to compete with experiment in quantities such as $g_A$.
This is an achievable goal if we combine our efforts.

\vspace{-0.4cm}
\section*{Acknowledgments}
\vspace{-0.3cm}

I would like to thank the following (listed in the order in which they communicated with me) for providing me with information and data:
K.-F.~Liu, M.~Engelhardt, R.~Horsley, X.~Feng, C.~Alexandrou, M.~Engelhard, S.~Sasaki, M.~Lin, T.~Rae, Y.~Yang, A.~Pochinsky, J.~Negele, M.~Sun, J.~Zanotti, H.~Wittig, J.~Green, E.~Shintani, B.~Menadue and B. Owen.
Thanks to the organizing committee for the opportunity to show the community these exciting (but sometimes overlooked) topics.
The speaker is supported by DOE grant DE-FG02-97ER4014.

\vspace{-0.4cm}
\bibliographystyle{apsrev}

\input{2012_latcon-proc.bbl}
\end{document}

%% file: intro.tex
Calculation of the low-energy hadronic matrix elements is important not only to make Standard Model predictions for comparison with experimental values
but also to provide the valuable inputs needed to assist in the searches for new physics. 
Flavor physics from lattice QCD has been playing an important role for the Standard Model inputs in heavy meson sector; B- and D-meson decay constants, B-mesons mixing parameters and more. For lighter mesons, such as pion and kaon systems, lattice QCD can determine many related quantities pretty reliably. 
However, when it comes to the nucleon, difficulties arise; this is frustrating, since unlike the mesons, which exist only ephemerally, nucleons are one of the building-block particles that compose us, and most measurements to understand Standard Model and probe new physics will involve interactions with nucleons. Given the difficulties and importance of nucleon quantities, in this review, I will focus on the nucleon, since there are still many challenges that need to be resolved and exciting new developments.

Nucleon matrix-element calculations on the lattice are harder in general because of the following problems. 
Firstly, we face a signal-to-noise problem. In Euclidean space the nucleon signal-to-noise ratio scales like $\exp(-M_N+3m_\pi/2)$ with Euclidean time; we need to extract the ground-state signal before noise becomes overwhelming.
The first excited-state $N(1440)$ is nearby; therefore, we have to be careful to distinguish the ground and excited signals to avoid contamination and to wait for larger time where the excited-state signal is relatively diminished, if necessary. However, this is at odds with the first problem; as we wait for larger time to avoid excited states, the signal-to-noise ratio decreases. Thus, we need high statistics and longer trajectories to have reliable numbers for experiments. 
Second, chiral perturbation theory (ChPT) is more difficult, because the $\Delta$ resonance is nearby. As a result, there are multiple expansions for ChPT, and some of them may have poor convergence, bringing the validity of chiral forms into question at the heaviest relevant pion mass. Thankfully, this issue may soon become moot in the physical pion-mass era. 
Also, larger volumes are required to avoid systematic uncertainties due to the finite volume. Low-statistics and heavy-pion studies of finite-volume effects cannot be directly applied to the case of light pion mass. And further, gauge ensembles are not always generated with nucleons in mind, resulting in suboptimally small volumes and low numbers of configurations.

Given the limited space, I will concentrate on reviewing a selection (colored by personal inclination) of exciting topics and progress since the last lattice conference plenary talk on the same topic. 
I start with hadron structure, focusing on new calculations of the gluonic momentum fraction and calculations of how pion structure is modified in a pion medium. 
Then, I will give two examples of calculations with applications to searches for physics beyond the Standard Model (BSM): the neutron electric dipole moment
and new tensor and scalar interactions in neutron $\beta$ decay. 
Lastly, I will review recent calculations of the nucleon axial charge and carefully consider its systematics. I will present more details on excited-state contamination, and I conclude that a better evaluation of the finite-volume effects is urgently needed to understand the current discrepancy between lattice calculations and the experimental value. 

%% file: struc.tex
The picture inside the hadron is usually described by studying longitudinal quark distributions in momentum space (structure functions, such as quark and gluon momentum fractions $\langle x^n \rangle_{q,g}$ to the $n^{\rm th}$ moment), or studying transverse quark distribution in coordinate space (through form factors).
However, the ambitious generalized parton distributions (GPDs) give a fully-correlated quark distribution in both coordinate and momentum space. 
Given the limited space, only some selected interesting updates will be presented here. Please also refer to the previous lattice hadron structure reviews~\cite{Alexandrou:2010cm,Renner:2010ks,Zanotti:2008zm,Hagler:2007hu,Orginos:2006zz} for omitted definitions.

\vspace{-0.2cm}
\subsection {Structure functions} 
\vspace{-0.2cm}
One of the fundamental questions in QCD is by what proportion the gluons and quarks make up the hadron and give their masses. The mass of pion and nucleon differ by an enormous amount despite there being only an increase of one quark in the hadron; the gluons must play a significant role. 
One way to directly study the question is to look at the gluon momentum fraction $\langle x \rangle_g$ (and its higher moments). 
Due to rotational symmetry breaking on the lattice, we can use either
${\cal O}_i^a=\tr ( \vec{\cal E} \times \vec{\cal B} )$
or
${\cal O}_i^b= \frac{2}{3}\tr ( \vec{\cal E}^2 +  \vec{\cal B}^2 )$
for $\langle x \rangle_g$ 
However, gluonic structure has been notoriously difficult to calculate with reasonable signals in lattice QCD, even for just the first moment. 
Despite these difficulties, the gluonic structure can be studied in the quenched approximation, and these new updates provide approaches and successful demonstrations that give some hope that the problem can be addressed.
Both $\chi$QCD~\cite{Liu:2012nz} and QCDSF~\cite{Horsley:2012pz} have made breakthroughs with updated studies of gluonic moments in the nucleon using Wilson gauge actions ($\beta=6.0$, $a^{-1}\approx 2$~GeV) with Wilson ($m_\pi \in \{480,650\}$) and nonperturbative $O(a)$-improved clover ($m_\pi \in \{480,650\}$) fermion actions, respectively.

$\chi$QCD~\cite{Liu:2012nz} uses a massless overlap Dirac operator for the gauge-field tensor as
$\tr_s (\sigma_{\mu\nu} D_{\rm ov}) \propto a^2 F_{\mu\nu}$
to compose the operator
${\cal O}_g = i \sum_k^3 F_{4k}F_{ki}$. 
The calculation is done using 500 gauge configurations, and two $Z_4$ noise sources to stochastically estimate $D_{\rm ov}$ for all space points at a given time. 
The same technique can be applied to look for general gluonic matrix-element contributions.

QCDSF uses the Feynman-Hellmann theorem by adding 
$$\beta \lambda \frac{1}{3} \left(\sum_{\vec{x},i} \re \tr_c \left[1-P_{i4}(\vec{x})\right] - \sum_{\vec{x},i<j} \re \tr_c \left[1-P_{ij}(\vec{x})\right] \right)$$
into the total QCD action. 
Then they calculate the nucleon mass (using standard two-point correlators, as in the nucleon sigma term) as a function of $\lambda$, which couples to the relevant gluonic operators; the derivative of the mass with respect to $\lambda$ contains the wanted gluonic matrix element:
$
\langle x \rangle_g = -\frac{2}{3} \frac{1}{M_N}\left.\frac{\partial M_N (\lambda)}{\partial \lambda}\right|_{\lambda=0}
$.

Both works use similar statistics, $O(500)$ configurations, and result in similar signal-to-noise ratio of the final result when extrapolated to the physical pion mass, about 15\%.
The QCDSF approach is interesting and computational cheap, but requires new configurations for each operator. One may need reliable reweighting techniques when moving to dynamical lattices. The $\chi$QCD approach is more robust; the same improved field operator can be used for any gluonic operator. However, it requires all-to-all overlap Dirac operators at any given time, so it is relatively expensive. However, one may be able to reuse $D_{\rm ov}$ for quark disconnected contributions and general gluonic operators.

Another interesting development is the study of how the hadron structure changes with in the presence of a medium. One of the most exciting topics in nuclear physics is the study of how hadronic properties are modified by the nuclear environment and how such modifications affect the properties of nuclei. In 1983, the European Muon Collaboration (EMC) reported surprising deep inelastic scattering (DIS) measurements of the structure function $F_2$ as a function of Bjorken $x$; the measurement  changed significantly between heavy nuclei and deuterium. 
Since then, many experiments have refined this measurement and looked at a variety of other isotopes and light nuclei. Many models were devised to try to explain the effect, and although many suggested factors appear to play some role, there is no universal understanding of the EMC effect yet. 
Such understanding would also be important to provide a better picture of how in-medium properties change to provide a baseline to constrain BSM physics, for example, in the NuTeV anomaly weak mixing angle experiment. This measurement lies 3 sigma away from the SM expectation, but the discrepancy may be explainable within QCD by environmental nucleon-effect corrections.

Unfortunately, multibaryon systems are complicated to calculate in LQCD; the noise increases, similarly to the sign problem. Therefore, we start with pion structure in a pion ($\pi^+$) medium. Multipion systems suffer from significant thermal effects due to the finite time extent of the box, due to pions going ``around the world'', and the problem worsens for light ensembles. To extract the energy of the multipion state, we need to not only consider the contribution of the $n$-$\pi$ system, but also all combinations of lower numbers of pions that can travel forward and backward (``around the world'') through the periodic time extent of the lattice volume.

\begin{figure}
\begin{center}
\includegraphics[width=0.52\columnwidth]{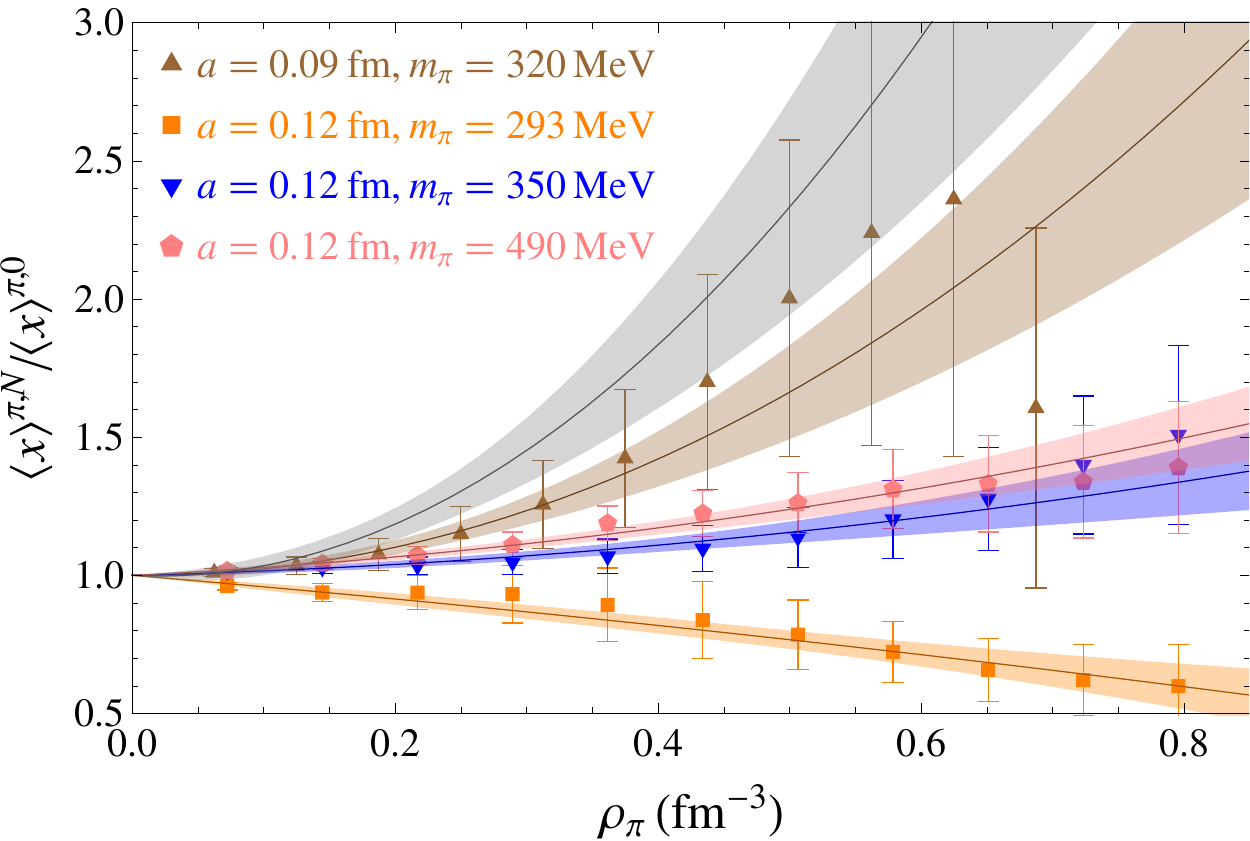}
\includegraphics[width=0.46\columnwidth]{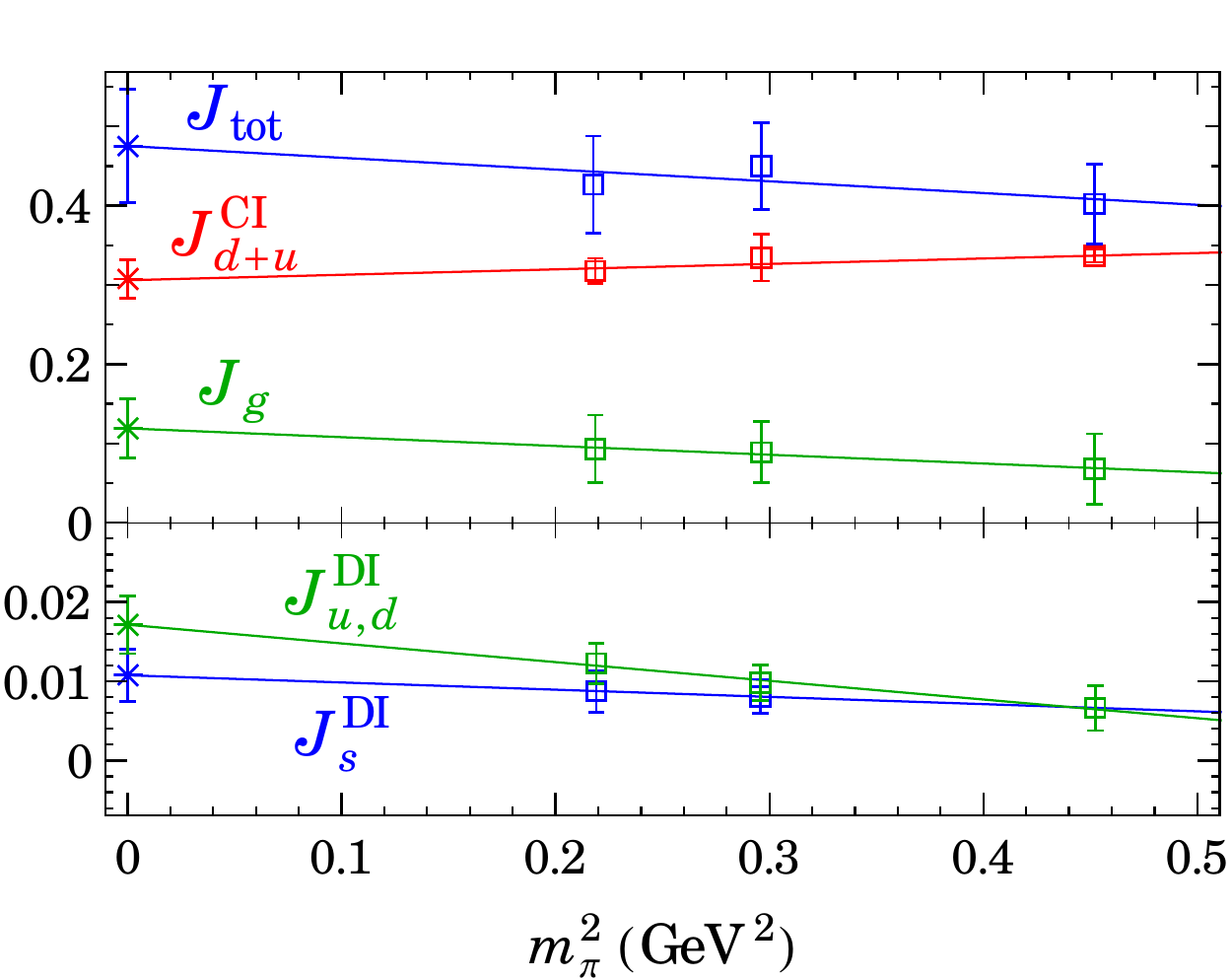}
\end{center}
\vspace{-0.5cm}
\caption{
(left) The pion quark momentum fraction in $\pi^+$, expressed as a ratio with its value in the vacuum,  $\langle x \rangle^{\pi,N}/\langle x \rangle^{\pi,0}$ as a function of pion density. 
(right) Total, gluon and connected up/down quark angular momentum to proton spin from $\chi$QCD collaboration.
Disconnected up/down and strange quark contributions are shown in the bottom panel.
}
\label{fig:emc-Js}
\end{figure}

This first lattice-QCD attempt to measure EMC effects for the pion momentum fraction in a pion medium uses pion masses ranging 290--490~MeV at 2 lattice spacings~\cite{Detmold:2011np}. On each ensemble, 3--5 source-sink separations are used to take the thermal-state degrees of freedom into account. The updated preliminary results for the ratio of the quark momentum fraction of the pion in medium versus the vacuum $\langle x \rangle^{\pi,N}/\langle x \rangle^{\pi,0}$ are shown on the
left-hand side of Fig.~\ref{fig:emc-Js} as function of pion density. The gray band is the preliminary extrapolated physical limit, indicating strong medium corrections to the first moment of the pion quark momentum fraction.

\vspace{-0.2cm}
\subsection {Form Factors}
\vspace{-0.2cm}
The low-$Q^2$ region of hadron form factors is a main focus of lattice calculations. We should devote more effort and resources to improving the lattice-QCD calculation of the charge radii; currently, the majority of calculations of the charge radii are about a factor of two smaller than the experimental value. Given the current tension and on-going theoretical and experimental efforts devoted to this quantity, we should seek better constraints on the radii.

Another interesting quantity is the induced pseudoscalar coupling constant $g_P^*$, which is important for muon-related physics; many of these experiments are designed to search for new physics. $g_P^*$ comes out as side product of the nucleon axial-current form factor, but in experiment, the induced form factor is defined at transfer momentum $0.88m_\mu^2$. Currently only RBC/UKQCD are reporting this number, and the constraints are poor at the moment. (See Fig.~5 in Ref.~\cite{Bhattacharya:2011qm} for a summary plot.) Recent experimental results from MuCap~\cite{Andreev:2012fj} have updated the $g_P^*$ precision to less than 1\%, 8.06(55).

\begin{figure}
\begin{center}
\includegraphics[width=0.48\columnwidth]{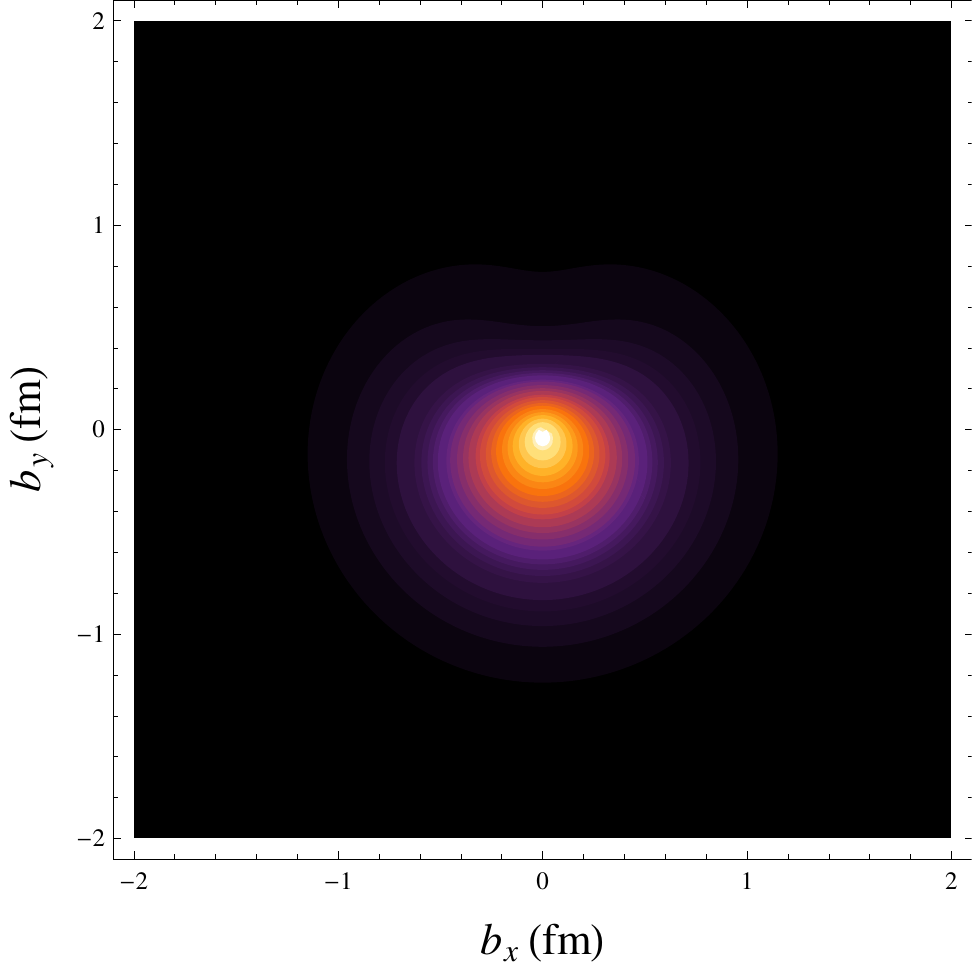}
\includegraphics[width=0.48\columnwidth]{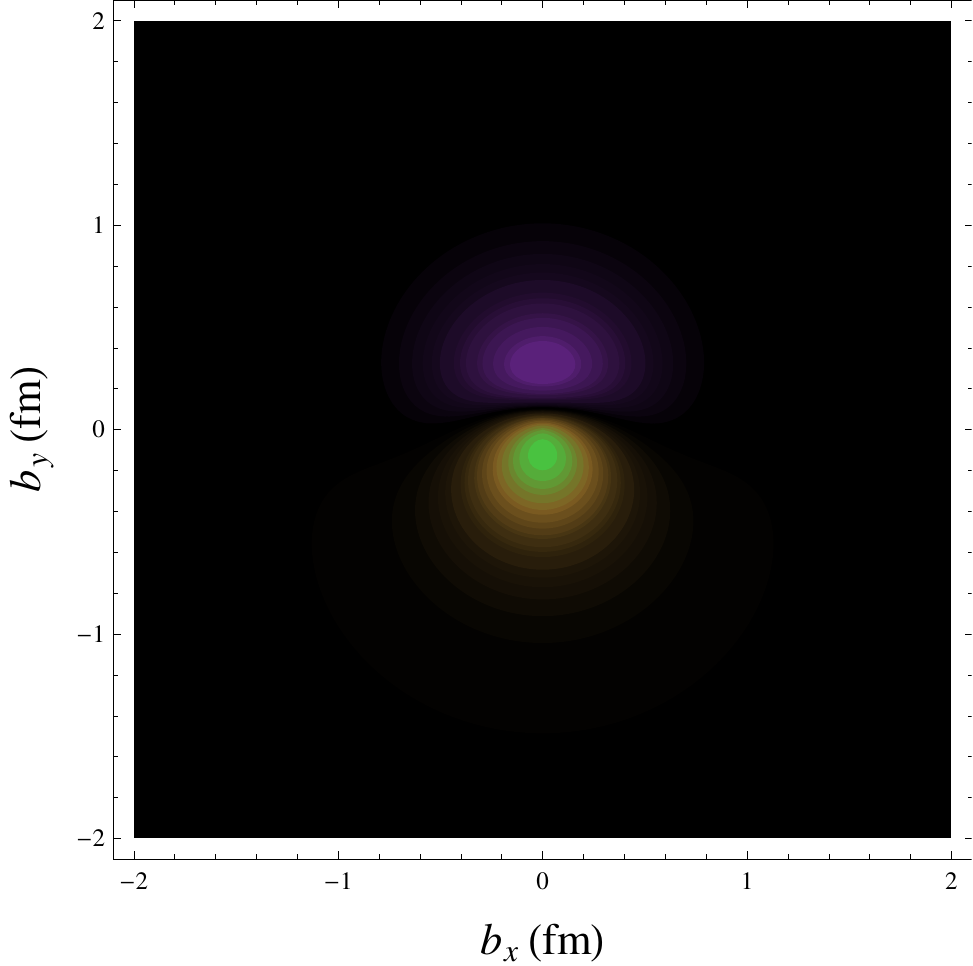}
\end{center}
\vspace{-0.5cm}
\caption{The transverse charge densities in a polarized proton (left) and neutron (right) with impact parameter $b_{x,y}$ in fm.
The color black indicates near-zero values, and purple, orange and white are increasingly positive (figures now in {\it Exploring the Heart of Matter} in the National Academies Press).
}
\label{fig:F-density}
\end{figure}

Studying the high-$Q^2$ region helps us investigate the distribution of charge density within the hadron. Even though at sufficiently large momenta, perturbative QCD should be able to describe the physics well, recent pion-to-two-photon data fail to converge to the asymptotic limit even at 40~GeV$^2$. An independent reliable nonperturbative QCD method would help. 
For example, work in Ref.~\cite{Lin:2010fv} extends the available $Q^2$ region by keeping nucleon operators that overlap both with zero-momentum and highly boosted ground-state nucleons. This helps out with the signal-to-noise at a given lattice spacing. 
Taking the large-$Q^2$ form-factor data from Ref.~\cite{Lin:2010fv}, we can map out the transverse charge-density distribution as a function of the impact distance $b$ in a polarized nucleon~\cite{Miller:2010nz,Carlson:2007xd,Miller:2007uy}:
\vspace{-0.2cm}
\begin{equation}
\label{eq:rho_T}
\rho_T(b) = \int_0^\infty\!\frac{Q\,dQ}{2\pi}J_0(bQ)F_1(Q^2) + \sin(\phi) \int_0^\infty\!\frac{Q^2\,dQ}{2\pi M_N}J_1(bQ)F_2(Q^2),
\end{equation} 
where $J_{0,1}$ are Bessel functions.
We can perform this integral numerically, using the lattice $F_{1,2}(Q^2)$ obtained by extrapolating the fit form to the physical pion mass.
Figure~\ref{fig:F-density} shows the results for the proton and neutron in the two-dimensional impact plane using lattice inputs. There are positive and negative charges surrounding the center, which in the neutron sum to zero. 
Ref.~\cite{Lin:2010fv} also show that using $Q^2 < 3\mbox{ GeV}^2$ will lead to significant deviations from the true short-distance behavior; compare with, for example, 6~GeV$^2$.
Note that the asymmetry in the distribution for a polarized nucleon is due to the relativistic effect of boosting the magnetic moment of the baryon. This induces an electric dipole moment that shifts the charge distribution.

\vspace{-0.2cm}
\subsection{Proton Spin}
\vspace{-0.2cm}
One of the fundamental questions in QCD is how the proton's spin $1/2$ is distributed among its constituents. The most naive intuition is that the three quarks carry the spin. Early experiments found the quark contribution makes up less than half of the total spin; this contradiction was dubbed the spin crisis, and many experiments and models attempted to address it. 
Now we better understand that significant contributions come from both intrinsic spin and orbital angular momentum. Ji~\cite{Ji:1995cu} gave a set of covariant operators for these that we can calculate on the lattice. 
QCDSF, LHPC and ETMC have been reported quark connected pieces for spin and orbital angular momentum for many years now (see reviews in Refs.~\cite{Alexandrou:2010cm,Renner:2010ks,Zanotti:2008zm,Hagler:2007hu,Orginos:2006zz,Lin:2011cr}). These results are renormalized in the $\overline{\rm MS}$ scheme at a scale of 2~GeV and shown on the right-hand side of Fig.~\ref{fig:emc-Js}. 
The extracted values are the stars on the left. As it turns out, the quarks contribute less than 50\% of the total nucleon spin; the total quark orbital angular momentum contribution is consistent with zero, so we must conclude that a majority of the proton spin comes from gluons, which is quite a surprise. 
The goal is still moving to lighter pion masses, addressing disconnected diagrams and evaluating gluon contributions that currently must be estimated from sum rules.

$\chi$QCD report the first lattice-QCD calculation including a complete set of the disconnected contributions and gluonic operators (see Ref.~\cite{Liu:2012nz} and private communication with Y.~Yang). On the right-hand side of Figure~\ref{fig:emc-Js}, we see their result as a function of pion mass squared. 
The proton spin breakdown is: quark spin contributed 50(2)\%, quark orbital angular momentum contributed 25(12)\% (mostly from the disconnected diagram, since they also see total zero quark contribution from the connected quark orbital angular momentum), and the total gluon spin contribution is 25(8)\%. 
However, the current calculation is still quenched Wilson action, and an update with overlap on DWF sea is in progress. It would be also interesting to see corroborating work from other collaborations using complete sets of the diagrams.

%% file: bsm.tex
There are many opportunities to probe physics beyond the Standard Model (BSM) with lattice-QCD calculations of hadronic matrix elements.
For example, we can calculate the leading hadronic vacuum contribution to the muon anomalous magnetic moment or evaluate processes involved in light-by-light contributions to it (by direct calculation, four-point correlators or meson-to-two photon form factors). The current discrepancy between the world-average theoretical and experimental values is around 3--4~$\sigma$; future improvement to the experimental data (reducing the error by a factor of 4 by 2015) will require further understanding and error-reduction on the nonperturbative QCD theoretical end. (See T.~Blum in these proceedings for a detailed review.) 
The nucleon scalar density, including its difficult disconnected piece, is an important low-energy constant needed as input for many BSM models of dark matter, such as supersymmetry (SuSy); R.~Young in these proceedings will also cover it. In this section, I will consider examples such as the neutron electric dipole moment (nEDM) and non-$\VmA$ interactions in neutron beta decay. 
Here I will briefly mention the progress made in lattice-QCD nEDM calculations and lattice-QCD calculations relevant to the detection of novel interactions in neutron beta decay.

\vspace{-0.2cm}
\subsection {Neutron Electric Dipole Moment (nEDM)}
\vspace{-0.2cm}
The neutron electric dipole moment is a measure of the distribution of positive and negative charge inside the neutron. To generate a finite nEDM, one needs processes that violate CP-symmetry, for example by adding a CP-odd $\theta$-term to the Lagrangian. The SM value of this quantity is very small, $10^{-30}$~$e\cdot\mbox{cm}$. Although experiments do not have the necessary precision to measure the SM value; many BSM models which predict values that are higher than the experimental upper bounds have been ruled out. This includes some parts of the parameter space for certain SuSy models. 
The most common type of lattice-QCD calculation mainly focuses on the leading $\theta$ term, the CP-odd contribution; a current combined analysis gives $O(30\%)$ in statistical error alone (see Fig.~1 in Ref.~\cite{Lin:2011cr} for a summary plot).
The precision of the lattice-QCD calculation needs to be greatly improved;
all-mode averaging (AMA) has been proposed to highly improve the current statistics even at near-physical pion mass (see Ref.~\cite{Blum:2012uh} and E.~Shintani in these proceedings). 
It may also be possible to combine the nucleon matrix elements with higher-order operators, such as quark electric dipole moment (qEDM) and chromoelectric dipole moment (CEDM) operators, (see T.~Bhattacharya, these proceedings) using effective field theory and proposed LQCD calculations.

\vspace{-0.2cm}
\subsection {Non-$\VmA$ Interactions in Neutron Beta Decay}
\vspace{-0.2cm}
The measurement of non-Standard Model contributions to precision neutron (nuclear) beta-decay measurements would give hints of potential BSM particles at the TeV scale; if new particles exist, their fundamental high-scale interactions could appear at low energy in the neutron beta-decay Hamiltonian as new terms:
\vspace{-0.2cm}
\begin{equation}
H_{\rm eff} = G_F \left( J_{V-A}^{\rm lept} \times J_{V-A}^{\rm quark} + \sum_i \varepsilon_i^{\rm BSM} \hat{O}_i^{\rm lept} \times \hat{O}_i^{\rm quark} \right),
\end{equation}
where $G_F$ is the Fermi constant, $J_{V-A}$ is the left-handed current of the indicated particle, and the sum includes operators with novel chiral structure. 
The new operators will enter with the coefficients $\varepsilon$ that are related to the TeV scale of the particles, similar to how dimensionful Fermi constant gives hints to the masses of the W and Z bosons of the electroweak theory before they were discovered. In the context of the nucleon, the new operators of this Hamiltonian will introduce low-energy coupling constants:
$g_T = \langle n | \overline{u}\sigma_{\mu\nu} d | p \rangle$, $g_S = \langle n | \overline{u} d | p \rangle$, 
and lattice QCD is a perfect theoretical tool to determine these constants precisely. The search for BSM physics proceeds experimentally by measuring the Fierz interference term and neutrino asymmetry parameter of neutron differential decay rate. Upcoming experiments at LANL (UCNB and UCNb) and ORNL (Nab) plan to measure $O_{\rm BSM}$ to the $10^{-3}$ level by 2013 (and maybe after a few years improve to $10^{-4}$). This only requires $g_{S,T}$ to be known to 10--20\% (after summing all the systematic uncertainties) to be useful as a BSM probe. 
For more details about experimental and theoretical work on this subject, refer to Ref.~\cite{Bhattacharya:2011qm} and R.~Gupta in these proceedings.

\begin{figure}
\begin{center}
\includegraphics[width=.48\textwidth]{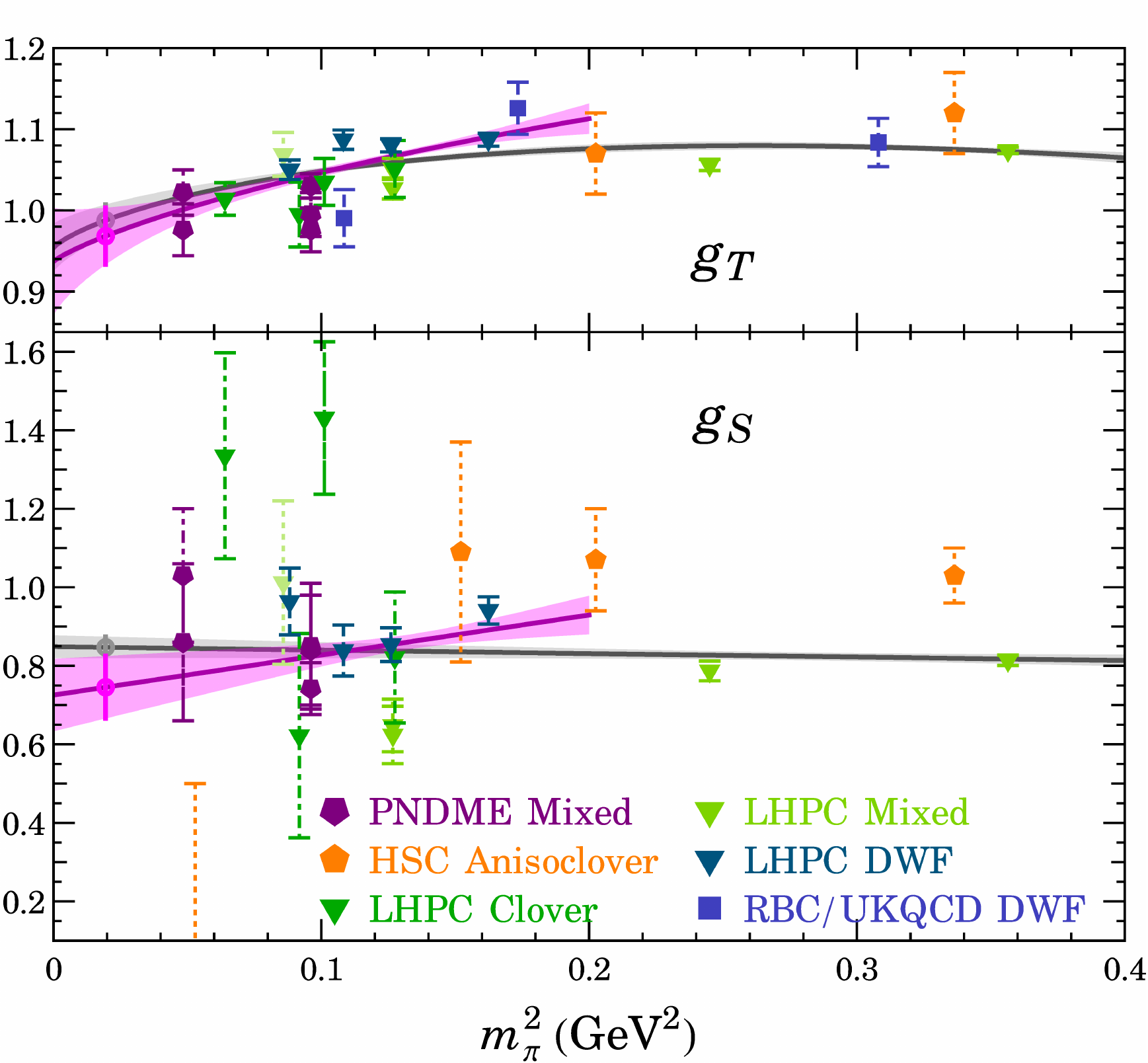}
\includegraphics[width=.48\textwidth]{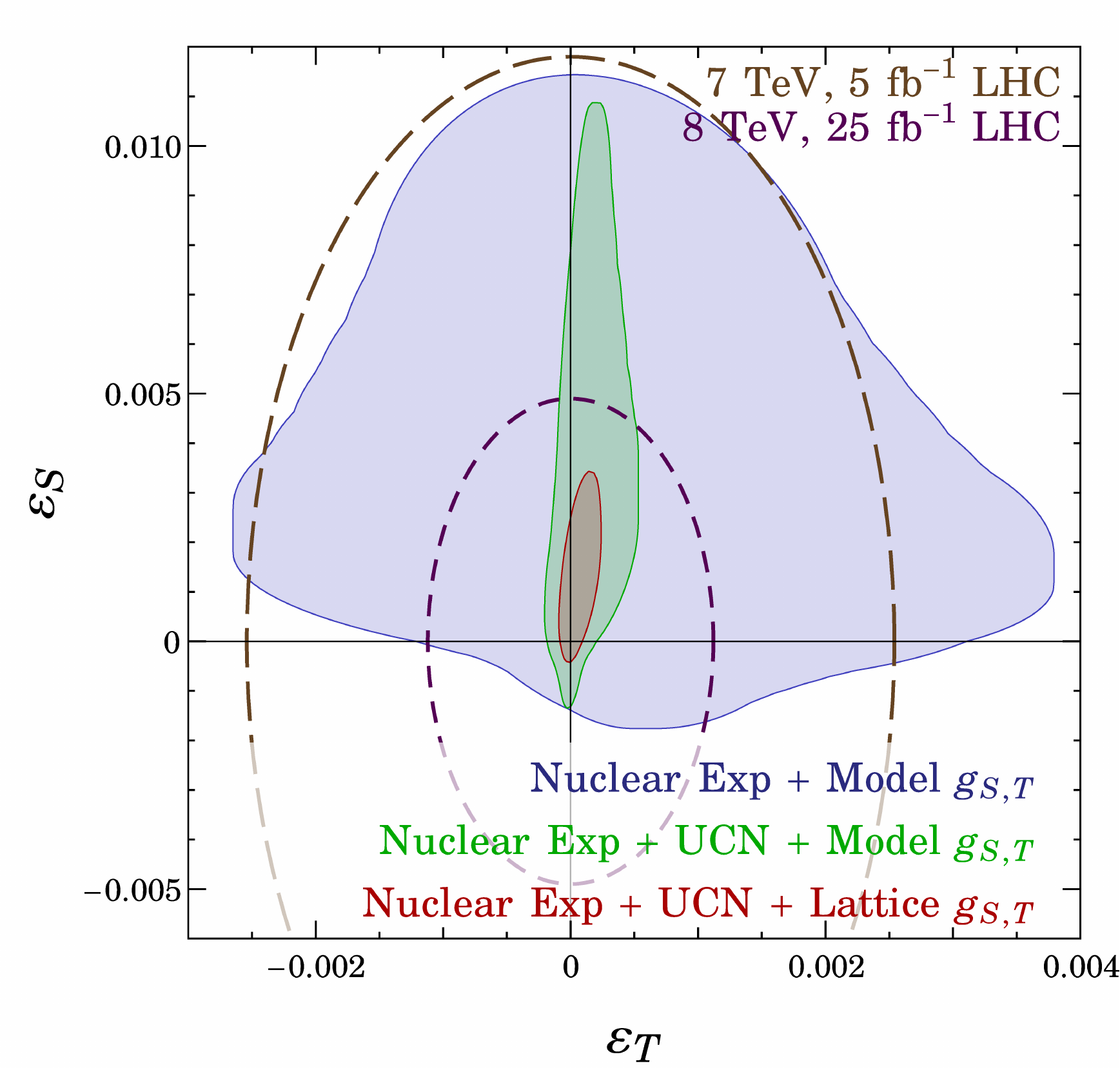}
\end{center}
\vspace{-0.5cm}
\caption{\label{fig:gTS-eSeT}
Global analysis of all $N_f=2+1(+1)$ lattice calculations of $g_T$ (upper-left) and $g_S$ (upper-right) with $m_\pi L >4$ and  $m_\pi T > 7$ cut to avoid systematics due to small spatial or temporal extent. The leftmost points are the extrapolated values at the physical pion mass. The two bands show extrapolations with different upper pion-mass cuts: $m_\pi^2 <0.4$ and  $m_\pi^2 <0.2$. The $m_\pi L<4$ data points are marked faded within each calculation; the lattice-spacings for each point are denoted by a solid line for $a\leq 0.06$~fm, dashed: $0.06 < a \leq 0.09$~fm, dot-dashed: $0.09 < a \leq 0.12$~fm and dotted: $a > 0.12$~fm. 
(Right) $\epsilon_{S}$-$\epsilon_{T}$ allowed parameter region using different experimental and theoretical inputs. The green and purple dashed lines are the constraints from first LHC run bound and near-term expectations with running to 2~GeV to compare with low-energy experiment.
}
\end{figure}

Ref.~\cite{Lin:2011cr} contains a summary of previous knowledge of these charges, experiments, models and previous lattice-QCD inputs, as applicable. Here we will concentrate on the updates available this year. PNDME (by Gupta) and LHPC (by J.~Green) reported updated calculations of $g_{S,T}$ after Ref.~\cite{Bhattacharya:2011qm}. 
We use the chiral formulation given in Refs.~\cite{Detmold:2002nf} and \cite{Green:2012ej} for tensor and scalar charges, respectively, to extrapolate to the physical pion mass (see the left-hand side of Fig.~\ref{fig:gTS-eSeT}). We see that the PNDME points greatly constrain the uncertainty due to chiral extrapolation in both cases and obtain $g_T^{\rm LQCD}= 0.978(35)$ and $g_S^{\rm LQCD}= 0.796(79)$ (statistical errors for now). 
The LQCD values are better determined than other theoretical estimations from various models, which give rather loose bounds on these quantities; for example, $g_S$ is estimated to be between 0.25 and 1.

These low-energy matrix elements, the tensor and scalar charges, can be combined with experimental data to determine the allowed region of parameter space for scalar and tensor BSM couplings (denoted $\epsilon$). Using the $g_{S,T}$ from the model estimations and combining with the existing nuclear experimental data%
\footnote{For example, nuclear beta decay $0^+ \rightarrow 0^+$ transitions and other processes, such as
$\beta$ asymmetry in Gamow-Teller $^{60}\mbox{Co}$,
longitudinal polarization ratio between Fermi and Gamow-Teller transitions in $^{114}\mbox{In}$,
positron polarization in polarized $^{107}\mbox{In}$
and beta-neutrino correlation parameters in nuclear transitions.}%
, we get the constraints shown as the outermost band on the left-hand side of Fig.~\ref{fig:gTS-eSeT}. Combining the expected experimental data in 2013 and existing measurements, and again, using the model inputs of $g_{S,T}$, we see the uncertainties in $\epsilon_{S,T}$ are significantly improved.
Finally, using our present lattice-QCD values of the scalar and tensor charges, combined with the expected 2013 precision of experimental bounds on deviation of these neutron-decay parameters from their SM values,
we found the constraints on $\epsilon_{S,T}$ are further improved, shown as the innermost region.
These upper bounds on the effective couplings $\epsilon_{S,T}$ will correspond to lower bounds for the scales $\Lambda_{S,T}$ at 5.6 and 10~TeV, respectively, for new physics in these channels.

How do the constraints from high-energy experiments compare? As demonstrated in Ref.~\cite{Bhattacharya:2011qm}, neither CDF nor D0 is sufficient to provide useful constraints.
We can estimate the $\epsilon_{S,T}$ constraints from LHC current bounds and near-term expectations through effective Lagrangian 
\begin{equation}
{\cal L} = -\frac{\eta_S}{\Lambda_S^2}V_{ud}(\overline{u}d)(\overline{e}P_L\nu_e)-\frac{\eta_T}{\Lambda_T^2}V_{ud}(\overline{u}\sigma^{\mu\nu}d)(\overline{e}\sigma_{\mu\nu}P_L\nu_e),
\end{equation} 
where $\eta_{S,T}=\pm 1$. 
The high-energy bound are scaled down to 2 GeV to compare with low-energy prediction;
more details can be found in Ref.~\cite{Bhattacharya:2011qm}.
By looking at events with high transverse mass from CMS and ATLAS in the $e\nu+X$ channel and comparing with the SM $W$ background, we estimated 90\%-C.L. constraints on $\eta_{S,T}$ based on the first LHC run, $\sqrt{s}=7$~TeV $L=10\mbox{ fb}^{-1}$ (the green line) and for a near-future run $\sqrt{s}=8$~TeV $L=25\mbox{ fb}^{-1}$ (the purple dashed line) on the right-hand side of Fig.~\ref{fig:gTS-eSeT}.

Furthermore, for detection experiments that have not yet observed any events such as $n$-$\bar{n}$ oscillations or proton decay, 
lattice-QCD calculations can provide low-energy constants to constrain the experimental search ranges.
The potential to search for new physics using these precision nucleon matrix elements during the LHC era and in advance of Project-X experiments at FNAL make lattice-QCD calculation of nucleon structure particularly timely and important.

%% file: gA.tex

\begin{figure}
\begin{center}
\includegraphics[width=0.32\columnwidth]{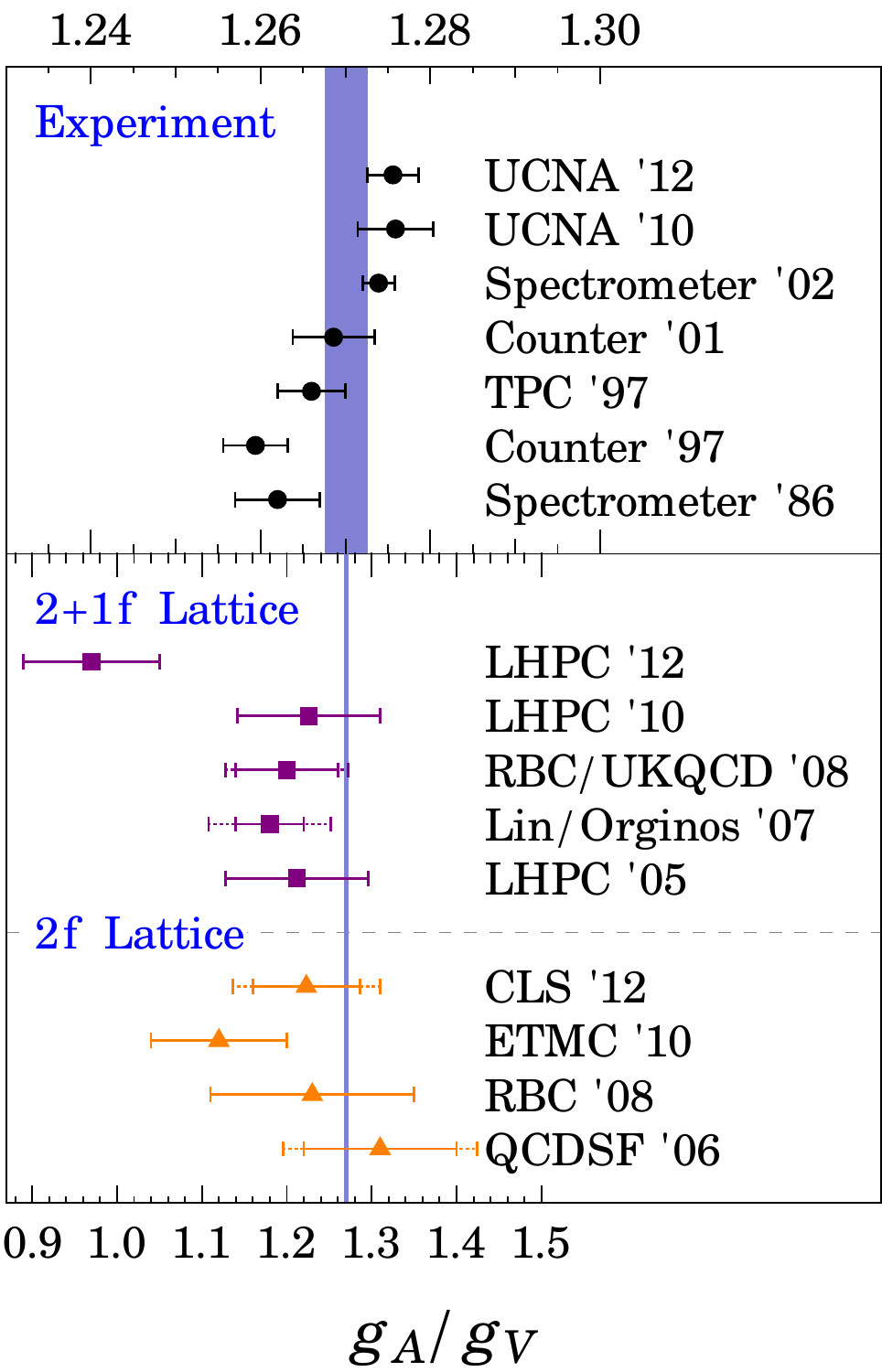}
\includegraphics[width=0.64\columnwidth]{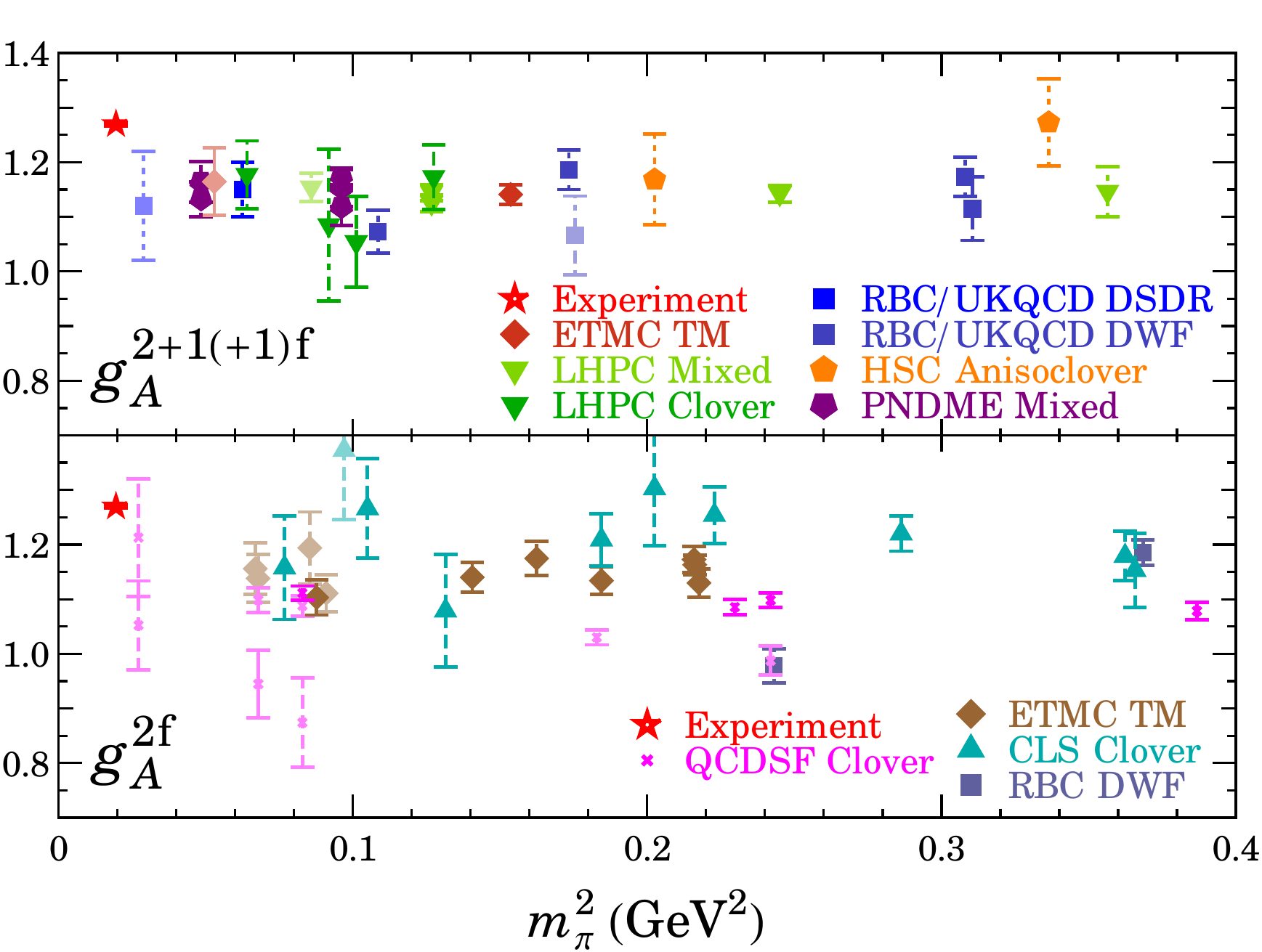}
\end{center}
\vspace{-0.5cm}
\caption{
(Left) Collected experimental values used in PDG 2012 average (the band) and the latest UCNA (2012) measurements on $g_A$; there has been a slow increase in $g_A/g_V$ in the past 15 years. The lower panel shows the physical-pion extrapolated $g_A$ numbers collected from dynamical 2+1-flavor and 2-flavor lattice calculations using $O(a)$-improved fermions~\cite{Khan:2006de,Lin:2008uz,Alexandrou:2010hf,Capitani:2012gj,Edwards:2005ym,Lin:2007ap,Yamazaki:2008py,Bratt:2010jn,Green:2012ud}. Note the change in scale and that most of the errorbars here are statistical only; a few calculations quoted systematics, which are added to statistical ones as outer errorbar bands, marked with dashed lines. (right) Published and ongoing $g_A$ calculations using $O(a)$-improved dynamical fermions, plotted as a function of $m_\pi^2$. The lattice spacings are grouped as described in the captions of Fig.~\protect\ref{fig:gTS-eSeT}. LHPC clover $m_\pi T \leq 7$ data points are excluded here.
}
\label{fig:gA}
\end{figure}

The nucleon axial charge $g_A$ is a very important fundamental measure of nucleon structure, with applications such as dictating the rate of proton-proton fusion, which is the first step in the thermonuclear reaction chains that power low-mass hydrogen-burning stars like the Sun, being an important input to measurements of BSM physics, affecting the extraction of $V_{ud}$ and neutrinoless double-beta decay.
Presently, $g_A$ is best known from the experimental measurement of neutron beta decay using polarized ultracold neutrons by UCNA collaboration, which dominates the PDG average with uncertainty at the 0.2\% level~\cite{PDG2012}. (See the upper panel of the left-hand side of Figure~\ref{fig:gA} for the collected experimental $g_A$ measurements used in PDG2012 and the recently updated UCNA number~\cite{Mendenhall:2012tz}.) This uses some assumptions about new few-TeV particles that can contribute to the measured process.
A precision lattice-QCD calculation can provide a cleaner $g_A$ directly from the SM and confront the experimental measurement, impacting many small-signal searches for SM-suppressed processes, as well as the astrophysical field for improving Solar-model physics.

In early nucleon matrix element calculations, $g_A$ was considered the low-hanging fruit in lattice nucleon matrix-element calculations, the gold-plated equivalent of say $f_\pi$ or $I=2$ $\pi$-$\pi$ scattering. The nucleon axial charge has been considered a benchmark quantity for lattice-QCD nucleon matrix elements.
Due to limited resources, we were not able to provide sufficient statistics nor close-to-physical pion masses to better quantify our systematics; as a result, the overall uncertainty is quite large and overlaps the entire experimentally allowed range.
However, many recent calculations, involving multiple lattice spacings (i.e. continuum extrapolated) and improved statistics have been able to close the gap.
Even as more computational resources become available, the best $g_A$ calculation is at 5\% statistical error and a few standard deviations away from the experimental values.
However, as seen on the right-hand side of Fig.~\ref{fig:gA}, these data points are 10--15\% lower than the experimental values, and those that seem to agree with experiment within 1--2~$\sigma$ tend to have large errorbars. A more reliable estimation of systematic uncertainties has not yet been made for the nucleon matrix elements. For example, estimates of the finite-volume effects might be too small.
Before we jump to a conclusion concerning the existence of new physics or contemplate inherent problems with lattice QCD, we should reexamine all the systematics and push to control them at a reliable level of precision (say, percent-level or smaller).

In the following, we consider the usual suspects for the systematics of a lattice measurement:
renormalization, lattice discretization artifacts, excited-state contamination,  chiral behavior and finite-volume corrections.

\vspace{-0.2cm}
\subsection {Renormalization and lattice-discretization artifacts}
\vspace{-0.1cm}
It is important to note that for the clover or $O(a)$-improved Wilson fermions, we also need to $O(a)$-improve the axial current $A_\mu^R=Z_A (a_A+b_A am_q)(A_\mu + ac_A\partial_\mu P)$ in the $g_A$ calculation, where the $b_A$ term vanishes in the chiral limit. Otherwise, the systematics due to lattice discretization artifacts would come in at $O(\Lambda_{\rm QCD}a)\approx 30$\% (for $a=0.12$~fm and $\Lambda_{\rm QCD}\approx 0.5$~GeV). Mainz group has good control over the current and renormalization through Schr\"odinger-functional scheme, while LHPC does not. PNDME plans to make three lattice-spacing calculations to study and remove discretization effects.
The current-improvement $O(a)$ term, although suppressed for chiral fermions at finite $a$, should still be carefully considered as a potential residual chiral symmetry breaking; for example, in the case of domain-wall fermions, non-small $m_{\rm res}$ can give rise to significant $O(a)$ artifacts.
Note that when one composes ratios such as $g_A/f_\pi$, it has been found that the ratio has better agreement with experiment at given lattice spacing. It is not only because the $Z_A$ is canceled, but also the $O(a)$ systematics are reduced if there are remaining significant $O(a)$ artifacts in the currents.

Among different lattice fermion actions, there are no observed significant artifacts on $g_A$ with lattice spacing below 0.12~fm. (See the right-hand side of Fig.~\ref{fig:gA} for lattice-spacing dependence. Data points are marked by different types of dashing for lattice spacings in the regions of 0.12, 0.09, 0.06~fm.)
It is safe to say that this is not the problem that causes the discrepancy from experiment (crisis) that we have observed, but it may become important and significant in the future when we try to become competitive with experiments to provide high-precision neutron-decay constants.
More resources should be devoted to increasing the statistics, at least to the $O(1\%)$ level, and we should not forget to revisit this systematic when higher-statistics measurements become available.

\vspace{-0.2cm}
\subsection {Chiral extrapolation}
\vspace{-0.2cm}
There is always a debate as to how reliable chiral effective theory (ChPT) is and how well it describes the data; there are diverse observations and nearly as diverse conclusions.
Statistics matters when it comes to whether a lattice-QCD has successfully provide a reliable $g_A$ or not.
In many of the chiral extrapolation shown on the left-hand side of Fig.~\ref{fig:gA}, using either heavy-baryon chiral perturbation theory (HBChPT) or small-scale expansion (SSE) formulation, one needs to set a number of low-energy constants; those constants could be biased tuned to match the experimental values.
We may need to consider using blinded techniques for extrapolation to avoid human bias in addition to making precise measurements and quoting systematics.
Luckily, calculations on ensembles at the physical pion mass are an attainable near-term goal; however, they will require large-enough volumes for a $g_A$ calculation and long trajectories.

\vspace{-0.2cm}
\subsection {Excited-state contamination}
\vspace{-0.2cm}
\begin{figure}
\begin{center}
\includegraphics[width=0.44\columnwidth]{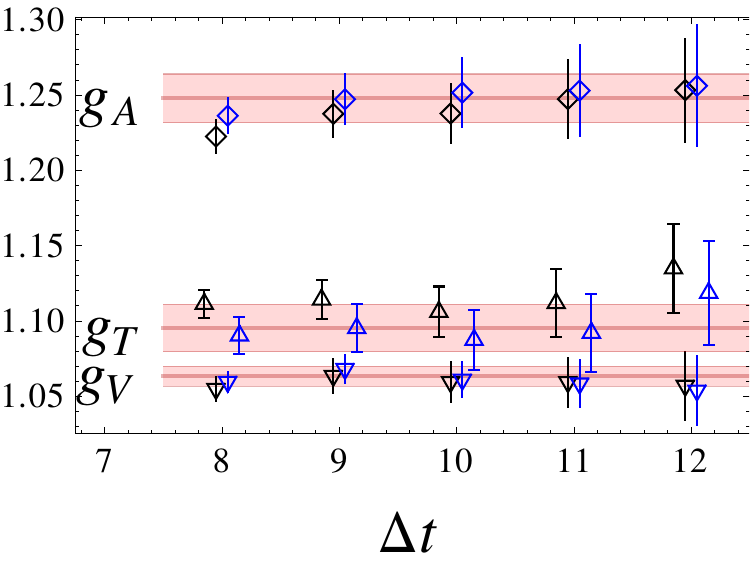}
\includegraphics[width=0.44\columnwidth]{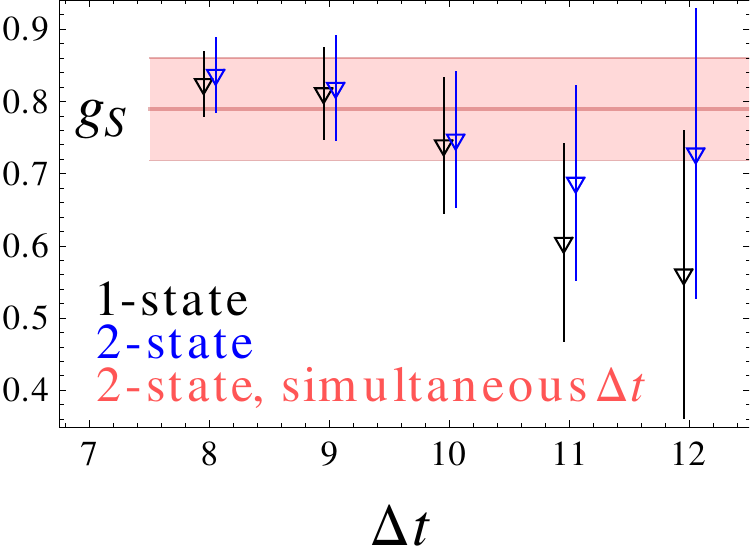}
\end{center}
\vspace{-0.5cm}
\caption{
The bare nucleon charges as functions of source-sink separation ($\Delta t \in [0.96,1.44]$~fm) by PNDME ($a \approx 0.12$~fm, $m_\pi \approx 310$~MeV) with $O(4000)$ measurements. The black (blue) points are the charge values obtained from one-state (two-state) fits at each $\Delta t$, while the band indicates a simultaneous fit to all $\Delta t$. Although the one- and two-state analyses give charges consistent within the statistical error, the two-state have smaller fluctuations about the central values and are more consistent with the simultaneous-fit analysis.
}
\label{fig:gVAST-dt}
\end{figure}
Excited-state contamination poses a problem due to the nearby Roper resonance on top of the nucleon's poor signal-to-noise problem at large Euclidean time, which scales like $e^{-(M_N-3m_\pi/2)t}$. In terms of three-point correlators, from which we extract $g_A$, to be confident that the signal is free of excited-state contamination, we need large source-sink separation; this large $t$ also worsens the noise. The usual recipe is to check the source-sink dependence at low statistics and identify one value for full production; however, it is difficult to be completely sure that the chosen value of the separation would be clean of excited-state contamination at the full statistical precision.

Strategies for dealing with the problem can be categorized in 3 main directions:
First, even if we limit ourselves to single-state analysis, we can simply increase the source-sink distance (say to 1.4 or 1.5~fm), where we can extract our favorite plateau with most of the kinematic and overlap factors canceling out, as done by RBC/UKQCD.
Alternatively, we can include the excited-state degrees of freedom and extract the ground-state nucleon by fitting. This was first proposed and demonstrated in Ref.~\cite{Lin:2008gv} in 2008. To facilitate accurate extraction of the excited state to disentangle it from the ground state, we can either add operators that overlap differently with the ground and excited states, or have high statistics to stabilize the analysis of multiple degrees of freedom. For example, see work done in Refs.~\cite{Lin:2008gv,Lin:2010fv}. More groups have started to adopt similar ideas into their analyses, including LHPC, CLS/Mainz and ETMC.

A recently proposed method called the ``summation'' method was introduced to take advantage of small-separation three-point correlators, extending the source-sink separation to where cleaner signal can be easily obtained given the same statistics. CLS/Mainz claim that the summation analysis resolves the discrepancy in smaller-plateau single-state analysis; however, although Mainz's studies observe a noticeable upward shift in the central values, their final numbers are consistent with a single-plateau analysis at 1.1~fm or so. LHPC makes a similar observation; their final ``summation'' analysis is consistent with the single-state analysis with largest separation. The method does not include both the transition matrix elements; when one works with nonzero momentum transfer, depending on the chosen kinematics, one should adopt a formulation that can pick up the dominant transition matrix and time factors so that excited-state transition form factors can be removed from the ground-state signal.

It is worth mentioning that ETMC uses high statistics with multiple source-sink separations in the best study of excited-state contamination for nucleon matrix elements~\cite{Alexandrou:2011aa}. They increased the statistics at the large source-sink separation by an order of magnitude so that all the matrix elements have comparable statistical errors.
ETMC also compared different analysis approaches, including summation (seven source-sink separations) and generalized eigenvector problem (GEVP), and found $g_A$ to have mild source-sink separation dependence as long as the nucleon operator is highly optimized to overlap with the ground-state; however, other matrix elements, such as the leading-moment of the quark momentum fraction, may not. Different quantities must be studied on a case-by-case basis.
CSSM adopted the variational method on matrix-element correlators for the meson case; future work with nucleon matrix elements is in progress.

To resolve the problem of excited-state contamination once and for all, the solution (in the author's opinion) would be to fully include the excited-state degrees of freedom (including 2 nucleon-Roper transition matrix elements and even the Roper-Roper one) and improve the statistics to stabilize the analysis.
An example is shown in Fig.~\ref{fig:gVAST-dt} with preliminary PNDME measurements at $O(4{\rm k})$ statistics; the 2-state analyses remain consistent even at smaller separation.
One might be able to push to smaller separation (but it may depend on the smearing and operator choices); at some point, the second-excited state and yet higher states must come in. One still needs to proceed with caution and not too aggressively reduce the separation.
RBC/UKQCD are adopting all mode averaging (AMA), and $\chi$QCD and JLQCD are using all-to-all approaches. Of course, an all-to-all propagator would be very useful for constructing an excellent approximation to the nucleon ground state, as in variational methods.

\vspace{-0.2cm}
\subsection {Finite-volume corrections}
\begin{figure}
\begin{center}
\includegraphics[width=0.43\columnwidth]{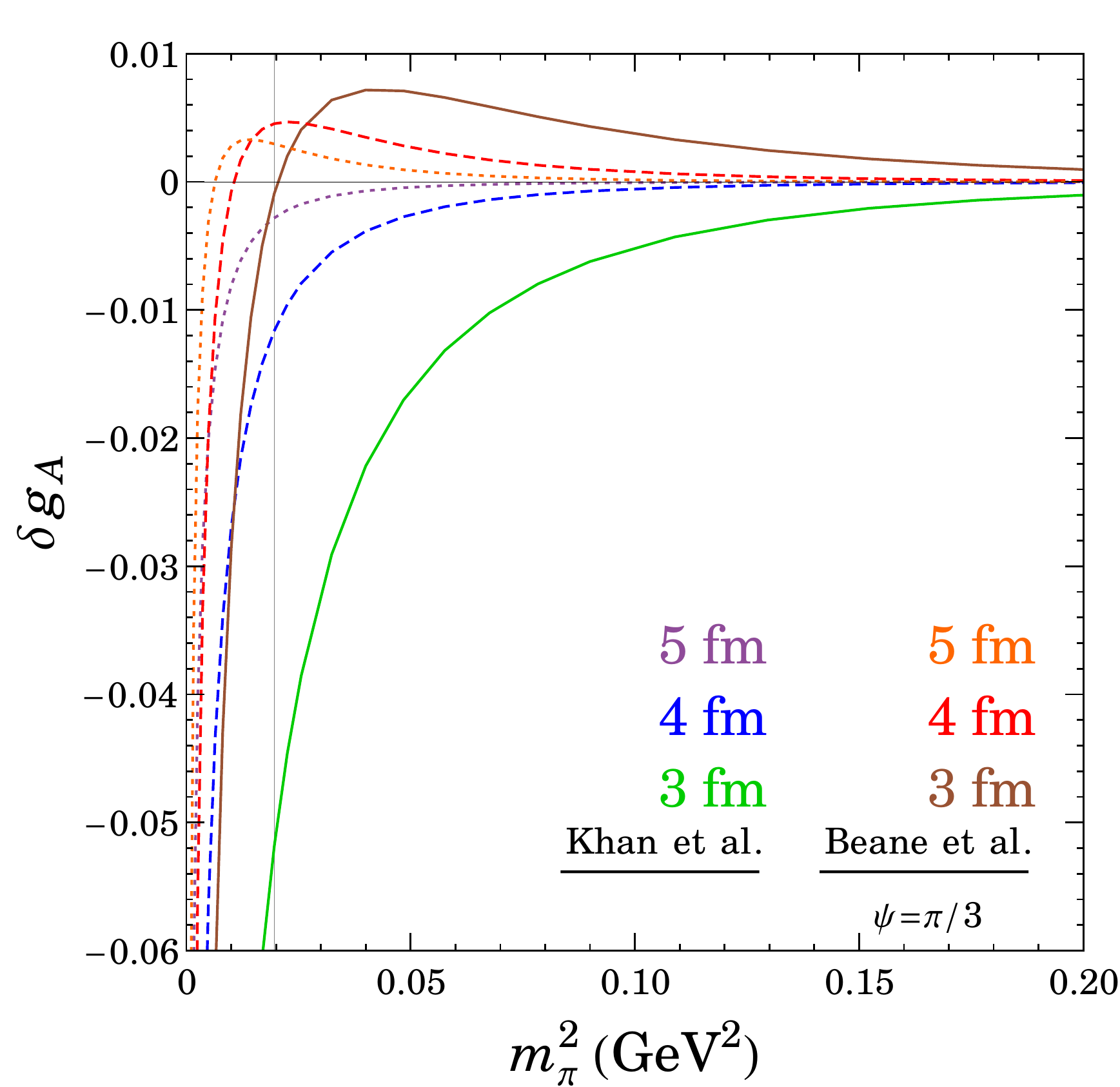}
\includegraphics[width=0.55\columnwidth]{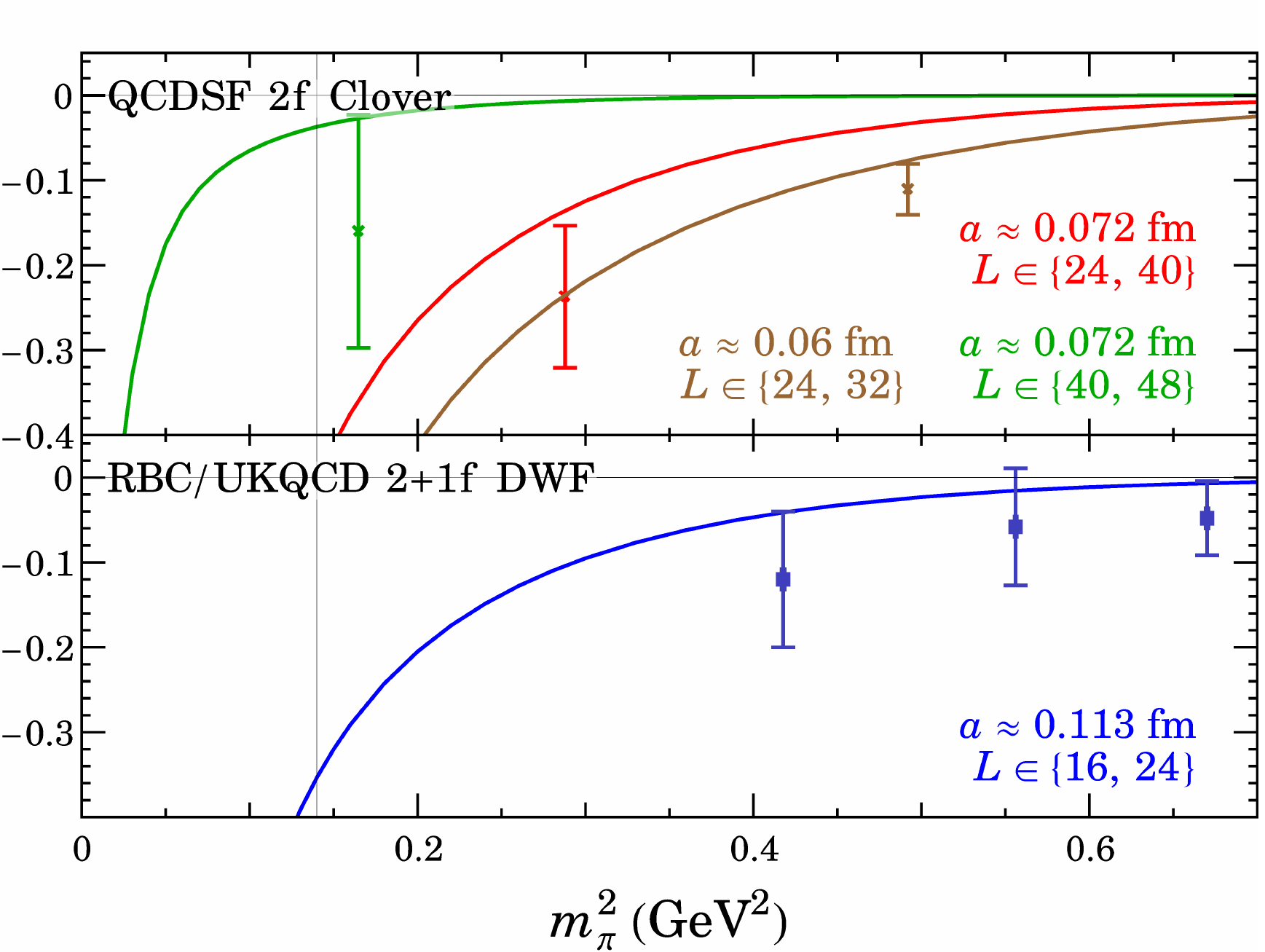}
\end{center}
\vspace{-0.5cm}
\caption{
(left) The finite-volume corrections for $g_A$ as a function of $m_\pi^2$ with the parameter choices set in Refs.~\cite{Beane:2004rf,Khan:2006de} at volume set to 3, 4, 5~fm. Note that the $\psi$ has changed to $\pi/3$ to better match with experimental coupling values. The formulation suggests finite volume correction at physical pion mass for 4~fm box is less than 1\%.
(right) The changes in $g_A$ as volumes changes at a fixed lattice spacing and various pion masses. The top panel selects examples of the QCDSF 2-flavor results, while the bottom one shows sample data from RBC/UKQCD. The lines indicate $g_A$ shifts corresponding to parameter choices in Ref.~\cite{Khan:2006de}, which indicate larger finite-volume corrections. Note that although the lattice values agree within 1--2 sigma with Ref.~\cite{Khan:2006de}, the central values are greater by up to an order of magnitude at the lightest QCDSF point or a factor of 2 for RBC/UKQCD. The volume corrections can cause up to 0.3 shift in these cases, which could account for the lattice-experimental discrepancy; we need to use high-statistics data to better understand this systematic.
}
\label{fig:FV}
\end{figure}

\vspace{-0.2cm}
The systematics caused by the finite-volume effects have been known to decrease the value of $g_A$ by significant amount.
Two commonly used formulations to determine finite-volume corrections are HBChPT~\cite{Beane:2004rf} and its variation SSE~\cite{Khan:2006de}. The finite-volume corrections in these references are very similar under the replacement of $f=\sqrt{2} F_\pi$, $c_A=\sqrt{2}g_{\Delta N}$, $g_1=-g_{\Delta \Delta}$, near-physical $\Delta$ and $g_A^0$ in the chiral limit.
(A different formulation regarding finite-volume corrections by CSSM can be found in Ref.~\cite{Hall:2012qn}).
The left-hand side of Fig.~\ref{fig:FV} shows two different sets of parameter choices on multiple selected volumes $L \in [3,5]$~fm as function of pion mass.
These parameters are set in Ref.~\cite{Khan:2006de} using a variety of constraints to
$F_\pi=86.2$~MeV, $c_A=1.5$, $g_1=2.6$, $g_A^0=1.15$,
while Ref.~\cite{Beane:2004rf} use SU(6) relations to derive
$g_A=1+(2/3)\cos^2\psi$, $g_{\Delta N}=-2 \cos \psi$, $g_{\Delta\Delta}=-3$.
The finite-volume corrections at physical pion mass are 0--0.004 and 0.05--0.004 for volume changing from 3 to~5 fm (corresponding to $m_\pi L=2.1$--3.6), which would be well below the statistical error at the current level of lattice-calculation precision. Note that Ref.~\cite{Khan:2006de} has a larger volume correction mainly due to larger coupling in $c_A$ terms.
One can also replace the formula with lattice $m_\pi/f_\pi$, but this only results in a small difference.
Groups that estimate their finite-volume corrections using the Refs.~\cite{Beane:2004rf,Khan:2006de} have reported small systematics to less than 1\% for $m_\pi L \approx 3$.

However, an earlier quenched RBCK study~\cite{Sasaki:2003jh} saw more significant effects, as shown in Refs.~\cite{Beane:2004rf,Khan:2006de}.
Recent RBC/UKQCD $N_f=2+1$ studies also see a significant central-value shift of about 0.05 and 0.12 with pion masses 670 and 420~MeV when the volume changes from  2.74 to 1.82~fm (which corresponds to $m_\pi L$ going from 9.3 to 6.2 and 5.3 to 3.9, respectively).
QCDSF has multiple lattice spacings ($a\in[0.06,0.078]$~fm) with various pion masses and volumes, and also sees a bigger volume correction than suggested in Refs.~\cite{Beane:2004rf,Khan:2006de}.

To compare the finite-volume corrections from ChPT and lattice data, we take the difference between two lattice $g_A$ values at fixed pion mass and lattice spacing but with differing volumes, and compare with the change predicted according to Ref.~\cite{Khan:2006de}; the right-hand side of Fig.~\ref{fig:FV} selects a subset of lattice calculation points from RBC/UKQCD and QCDSF.
Note that although the lattice volume-shifts seems to be at the edges of their errorbars of the Ref.~\cite{Khan:2006de} line, they are systematically larger in magnitude with central values greater by a factor of 2 to 10.

Finite-volume corrections need to be done at smaller pion masses; many groups estimate the finite-volume correction at larger pion mass with the same $m_\pi L$, but we have seen that yet larger $m_\pi L$ are needed for smaller pion masses. This effect could cause underestimation of the finite-volume corrections in $g_A$.
The volume changes can cause shifts of up to 0.3 in the lattice $g_A$ as suggested from selected lattice data in Fig.~\ref{fig:FV}; this alone could explain the lattice and experiment discrepancy. We need to study these effects with greater statistics to rule out or better understand the form of the finite-volume effects.

